\documentclass[aps,prl,twocolumn,superscriptaddress,showpacs,preprintnumbers,amsmath,amssymb]{revtex4}

\usepackage{graphicx} 
\usepackage{dcolumn}  

\graphicspath{{figs/}}

\def\nbb{\mbox{$152$}}

\def\YIELDWOSYST{
  \mbox{$25.6^{+9.3}_{-8.4}$}}
\def\YIELD{
  \mbox{$25.6^{+9.3}_{-8.4} ({\rm stat}) ^{+1.6}_{-1.4} ({\rm syst})$}}
\def\SIGMAWORHOPI{\mbox{$3.5$}}
\def\SIGMA{\mbox{$3.4$}}

\def\BR{
  \mbox{$(1.7 \pm 0.6 ({\rm stat}) \pm 0.2 ({\rm syst}))\times 10^{-6}$}}

\def\BabarBR{
  \mbox{$(2.1 \pm 0.6 \pm 0.3) \times 10^{-6}$}}

\def\fb{\mbox{fb$^{-1}$}}
\def\bb{\mbox{$B\overline{B}$}}
\def\qq{\mbox{$q\overline{q}$}}
\def\mbc{\mbox{$M_{\rm bc}$}}
\def\de{\mbox{$\Delta E$}}

\def\bz{\mbox{$B^0$}}

\def\bp{\mbox{$B^+$}}

\def\dz{\mbox{$D^0$}}
\def\dzb{\mbox{$\overline{D}{}^0$}}

\def\kppim{\mbox{$K^+ \pi^-$}}

\def\pizpiz{\mbox{$\pi^0 \pi^0$}}

\def\dzbpip{\mbox{$\dzb(\to\kppim\pi^0)\pi^+$}}

\begin{document}

\preprint{\vbox{ \hbox{   }
}}

\title{\quad\\[0.5cm] \boldmath
  Evidence for $\bz\to\pizpiz$
}


\affiliation{Budker Institute of Nuclear Physics, Novosibirsk}
\affiliation{Chiba University, Chiba}
\affiliation{University of Cincinnati, Cincinnati, Ohio 45221}
\affiliation{University of Frankfurt, Frankfurt}
\affiliation{Gyeongsang National University, Chinju}
\affiliation{University of Hawaii, Honolulu, Hawaii 96822}
\affiliation{High Energy Accelerator Research Organization (KEK), Tsukuba}
\affiliation{Hiroshima Institute of Technology, Hiroshima}
\affiliation{Institute of High Energy Physics, Chinese Academy of Sciences, Beijing}
\affiliation{Institute of High Energy Physics, Vienna}
\affiliation{Institute for Theoretical and Experimental Physics, Moscow}
\affiliation{J. Stefan Institute, Ljubljana}
\affiliation{Kanagawa University, Yokohama}
\affiliation{Korea University, Seoul}
\affiliation{Kyungpook National University, Taegu}
\affiliation{Institut de Physique des Hautes \'Energies, Universit\'e de Lausanne, Lausanne}
\affiliation{University of Ljubljana, Ljubljana}
\affiliation{University of Maribor, Maribor}
\affiliation{University of Melbourne, Victoria}
\affiliation{Nagoya University, Nagoya}
\affiliation{Nara Women's University, Nara}
\affiliation{National Kaohsiung Normal University, Kaohsiung}
\affiliation{National Lien-Ho Institute of Technology, Miao Li}
\affiliation{Department of Physics, National Taiwan University, Taipei}
\affiliation{H. Niewodniczanski Institute of Nuclear Physics, Krakow}
\affiliation{Nihon Dental College, Niigata}
\affiliation{Niigata University, Niigata}
\affiliation{Osaka City University, Osaka}
\affiliation{Osaka University, Osaka}
\affiliation{Panjab University, Chandigarh}
\affiliation{Peking University, Beijing}
\affiliation{Princeton University, Princeton, New Jersey 08545}
\affiliation{RIKEN BNL Research Center, Upton, New York 11973}
\affiliation{University of Science and Technology of China, Hefei}
\affiliation{Seoul National University, Seoul}
\affiliation{Sungkyunkwan University, Suwon}
\affiliation{University of Sydney, Sydney NSW}
\affiliation{Tata Institute of Fundamental Research, Bombay}
\affiliation{Toho University, Funabashi}
\affiliation{Tohoku Gakuin University, Tagajo}
\affiliation{Tohoku University, Sendai}
\affiliation{Department of Physics, University of Tokyo, Tokyo}
\affiliation{Tokyo Institute of Technology, Tokyo}
\affiliation{Tokyo Metropolitan University, Tokyo}
\affiliation{Tokyo University of Agriculture and Technology, Tokyo}
\affiliation{University of Tsukuba, Tsukuba}
\affiliation{Utkal University, Bhubaneswer}
\affiliation{Virginia Polytechnic Institute and State University, Blacksburg, Virginia 24061}
\affiliation{Yokkaichi University, Yokkaichi}
\affiliation{Yonsei University, Seoul}
  \author{S.~H.~Lee}\affiliation{Seoul National University, Seoul} 
  \author{K.~Suzuki}\affiliation{High Energy Accelerator Research Organization (KEK), Tsukuba} 
  \author{K.~Abe}\affiliation{High Energy Accelerator Research Organization (KEK), Tsukuba} 
  \author{K.~Abe}\affiliation{Tohoku Gakuin University, Tagajo} 
  \author{T.~Abe}\affiliation{High Energy Accelerator Research Organization (KEK), Tsukuba} 
  \author{I.~Adachi}\affiliation{High Energy Accelerator Research Organization (KEK), Tsukuba} 
  \author{Byoung~Sup~Ahn}\affiliation{Korea University, Seoul} 
  \author{H.~Aihara}\affiliation{Department of Physics, University of Tokyo, Tokyo} 
  \author{K.~Akai}\affiliation{High Energy Accelerator Research Organization (KEK), Tsukuba} 
  \author{M.~Akatsu}\affiliation{Nagoya University, Nagoya} 
% ...
\collaboration{The Belle Collaboration}

\noaffiliation



\begin{abstract}
  We report evidence for the decay $\bz\to\pizpiz$.
  The analysis is based on a data sample of $\nbb$ million
  $\bb$ pairs collected at the $\Upsilon$(4S) resonance with
  the Belle detector at the KEKB $e^+e^-$ storage ring.
  We detect a signal for $\bz\to\pizpiz$
  with a significance of $\SIGMA$ standard deviations,
  and measure the branching fraction to be $\BR$.
\end{abstract}


\maketitle


{\renewcommand{\thefootnote}{\fnsymbol{footnote}}}

Recent measurements at $B$ factories have improved significantly 
our knowledge of CP violation and heavy flavour physics.
In particular, measurements of the mixing-induced CP violation
parameter $\rm{sin} 2\phi_1$~\cite{phi1_belle,phi1_babar}
lend strong support to the Kobayashi-Maskawa mechanism~\cite{km}.
It is of great importance to test precisely this theory 
with complementary measurements,
such as those of the other unitarity triangle angles
$\phi_2$ and $\phi_3$~\cite{pdg}.

The most promising technique for measuring $\phi_2$ is by studying
time dependent asymmetries in the $B^0 \to \pi^+\pi^-$ 
system~\cite{phi2_belle,phi2_babar}.
The extraction of $\phi_2$ from the observables is not trivial,
however, because there are contributions from more than one weak phase.
In order to disentangle $\phi_2$, 
an isospin analyses of the $\pi\pi$ system can be performed~\cite{isospin}.
One essential ingredient for these procedures, 
and hence for the measurement of $\phi_2$, is knowledge of the 
branching fraction for the decay $B^0 \to \pi^0\pi^0$.
Such knowledge would also play a pivotal role in the understanding
of charmless hadronic $B$ decays.
Previously, upper limits for the branching fraction of $B^0 \to \pi^0\pi^0$ 
of $(3-6) \times 10^{-6}$
have been reported~\cite{hh_belle,hh_babar,hh_cleo}.
Theoretical predictions are typically around or below
$1 \times 10^{-6}$~\cite{pi0pi0_predictions},
but phenomenological models incorporating large rescattering effects 
can accommodate larger values~\cite{theory}. 
The BaBar group recently measured this branching fraction to be $\BabarBR$~\cite{BaBar_new}.

In this paper, we report evidence for the decay $\bz\to\pizpiz$.
The results are based on a 140 $\fb$ ($\nbb$ M $\bb$ pairs) collected with the Belle detector at the KEKB $e^+e^-$ storage ring~\cite{kekb}.
KEKB operates at a center--of--mass (CM) energy of $\sqrt{s} = 10.58 \ {\rm GeV}$, corresponding to the mass of the $\Upsilon$(4S) resonance.
Throughout this paper, neutral and charged $B$ mesons are assumed
to be produced in equal amounts at the $\Upsilon$(4S),
and the inclusion of charge conjugate modes is implied.

The Belle detector is a large-solid-angle spectrometer consisting
of a three-layer silicon vertex detector, a 50-layer central
drift chamber (CDC), an array of threshold Cherenkov counters with
silica aerogel radiators (ACC), time-of-flight scintillation counters,
and an electromagnetic calorimeter comprised of CsI(Tl) crystals
(ECL) located inside a superconducting solenoid coil that provides a
1.5 T magnetic field.
An iron flux-return located outside of the coil is instrumented to
detect $K^0_L$ mesons and to identify muons.
A detailed description of the Belle detector can be found
elsewhere~\cite{belle}.

Pairs of photons with invariant masses in the range 
$115 \ {\rm MeV}/c^2 < m_{\gamma\gamma} < 152 \ {\rm MeV}/c^2$ are used
to form $\pi^0$ mesons; this
corresponds to a window of $\pm 2.5\sigma$ about the nominal $\pi^0$ mass,
where $\sigma$ denotes the experimental resolution, approximately $8 \ {\rm MeV}/c^2$.
The measured energy of each photon in the laboratory frame is
required to be greater than $50 \ {\rm MeV}$ in the barrel region, 
defined as $32^{\circ} < \theta_{\gamma} < 128^{\circ}$, and greater than
$100 \ {\rm MeV}$ in the end-cap regions,
defined as $17^{\circ} \le \theta_{\gamma} \le 32^{\circ}$ and 
$128^{\circ} \le \theta_{\gamma} \le 150^{\circ}$, 
where $\theta_{\gamma}$ denotes the polar angle of the photon with respect to the beam line.

Signal $B$ candidates are formed from pairs of $\pi^0$ mesons and 
are identified by their beam energy constrained
mass $\mbc = \sqrt{E_{\rm beam}^{*2} - p_B^{*2}}$ and energy difference
$\de = E_B^* - E_{\rm beam}^*$, 
where $E_{\rm beam}^*$ denotes the beam energy and 
$p_B^*$ and $E_B^*$ are the momentum and energy, respectively, of the reconstructed $B$ meson,
all evaluated in the $e^+e^-$ CM frame.
We require
$5.2 \ {\rm GeV}/c^2 < \mbc < 5.3 \ {\rm GeV}/c^2$ and 
$-0.3 \ {\rm GeV} < \de < 0.5 \ {\rm GeV}$.
The signal efficiency of the kinematic reconstruction is
estimated using GEANT-based~\cite{geant} Monte Carlo (MC) simulations
and found to be 26\%.
The resolution for signal is approximately $4 \ {\rm MeV}/c^2$ for $\mbc$ and $^{+ 50}_{-100} \ { \rm MeV}$ for $\de$.

We consider background from other $B$ decays and from 
$e^+e^-\to\qq$ ($q = u$, $d$, $s$, $c$) continuum processes.
A large generic MC sample shows that backgrounds
from $b\to c$ decays are negligible.
Among charmless $B$ decays, the only significant background is
$B^+ \to \rho^+\pi^0$.
We take these events, which populate the negative $\de$ region, into account in the signal extraction described below.

The dominant background is due to continuum processes. 
We discriminate signal events from the $\qq$ background 
using the event topology. 
In order to increase the expected sensitivity to the signal, 
we have improved the continuum rejection technique
used in our previous publication~\cite{hh_belle}.
Previously, we have defined modified Fox-Wolfram moments~\cite{fw}
that treat particles involved in the signal $B$ candidate ($s$) 
separately from those in the rest of the event ($o$).
We extend this idea, taking into account the missing momentum
in the event, which we treat as a third category ($m$).
We achieve some additional discrimination by considering 
charged and neutral particles in the $o$ category independently,
and by taking the correlations of charges into account.
In our previously used continuum rejection technique,
the moments are normalized relative to the zeroth moment;
in this improved technique we do not normalize in this way. 
We combine 16 modified moments with the scalar sum of the transverse 
momentum into a Fisher discriminant~\cite{fisher},
and tune coefficients separately for seven categories of missing mass squared to maximize the separation between signal and background.
MC studies indicate that this redefinition of the Fisher discriminant
leads to a $24\%$ improvement in the maximum value of the figure of merit (FOM) defined as
$N_s^{\rm exp}/\sqrt{N_{BG}^{\rm exp}}$,
where $N_s^{\rm exp}$ and $N_{BG}^{\rm exp}$
denote the expected signal and observed background yields in a region
$5.27 \ {\rm GeV}/c^2 < \mbc < 5.29 \ {\rm GeV}/c^2$ and
$-0.20 \ {\rm GeV} < \de < 0.05 \ {\rm GeV}$.

The angle of the $B$-meson flight direction with respect to the beam
axis ($\theta_B$) provides additional discrimination.
A likelihood ratio 
${\cal R}_s = {\cal L}_s / ({\cal L}_s + {\cal L}_{q\overline{q}})$ 
is used as the discrimination variable,
where ${\cal L}_s$ denotes the product of the individual Fisher and
$\theta_B$ likelihoods for the signal and 
${\cal L}_{q\overline{q}}$ is that for the $\qq$ background. 
The likelihood functions
are derived from MC for the signal and from 
events in the $\mbc$ sideband region
($5.20 \ {\rm GeV}/c^2 < \mbc < 5.26 \ {\rm GeV}/c^2$) 
for the $\qq$ background.
We find additional discrimination between signal and background
using the Belle standard algorithm for $b$-flavour tagging~\cite{phi1_belle,phi2_belle}.
The flavour tagging procedure yields two outputs:
$q = \pm 1$ (which we ignore), 
indicating the flavour of the other purported $B$ in the event,
and $r$, which takes values between 0 and 1 and is a measure of the confidence that the $q$ determination is correct.
Events with a high value of $r$
are considered well-tagged and are therefore
unlikely to have originated from continuum processes.
Moreover, we find that 
there is no strong correlation between $r$ and 
any of the topological variables used above to separate signal from continuum.

\begin{figure}
\resizebox{!}{0.35\textwidth}{\includegraphics{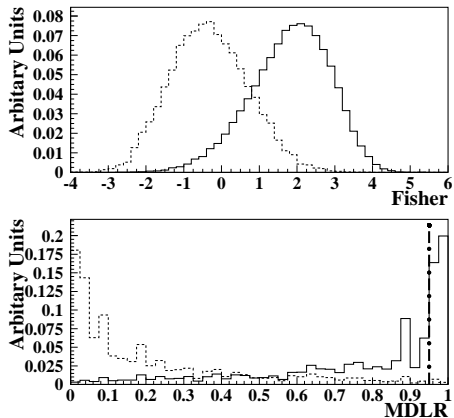}}
  \caption{
    \label{fig:mdlr}
    Distributions of variables used for continuum suppression.
      (Top) Fisher discriminant using modified Fox-Wolfram moments for signal MC (solid line) and for events in the continuum dominated sideband region (dashed line);
      (bottom) MDLR for signal MC (solid line) and sideband data (dashed line).
     The dot-dashed line at ${\rm MDLR = 0.95}$
     indicates the selection requirement made.
     Apparent structure in the MDLR distribution arises from the several
     $B$-tagging modes that contribute to the $r$ distribution.
  }
\end{figure}

We combine $r$ and ${\cal R}_s$ into a single 
multi-dimensional likelihood ratio (MDLR) defined as ${\cal L}^{\rm MDLR}_s / ({\cal L}^{\rm MDLR}_s + {\cal L}^{\rm MDLR}_{q\overline{q}})$, where ${\cal L}^{\rm MDLR}_s$ denotes the likelihood determined by the $r$--${\cal R}_s$ two-dimensional distribution for the signal and ${\cal L}^{\rm MDLR}_{q\overline{q}}$ is that for the $\qq$ background.
We then make a requirement on the likelihood ratio that maximizes the FOM.
Incorporating the flavour tagging information in this way gives a 4\% improvement in the FOM as compared to making a selection requirement on ${\cal R}_s$ alone.
This criterion eliminates 99\% of the $\qq$
background while retaining 39\% of the signal.
The MDLR distributions for signal MC and for sideband data are shown
in Fig.~\ref{fig:mdlr}. 
We verified that there is no correlation between the MDLR cut and $M_{bc}$ in the continuum MC events.

\begin{figure}
\resizebox{!}{0.35\textwidth}{\includegraphics{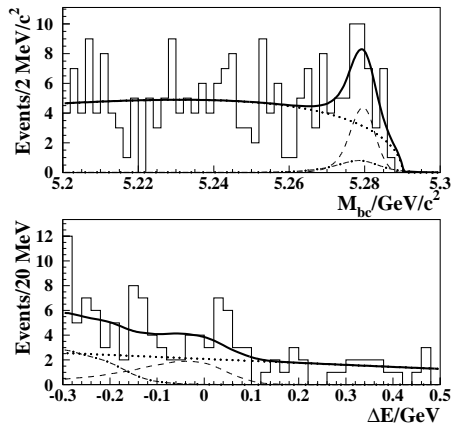}}
  \caption{
    \label{fig:fit_result}
    Result of the fit described in the text.
    (Top) $\mbc$ projection for events which satisfy
    $-0.2 \ {\rm GeV} < \de < 0.05 \ {\rm GeV}$;
    (bottom) $\de$ projection for events which satisfy
    $5.27 \ {\rm GeV}/c^2 < \mbc < 5.29 \ {\rm GeV}/c^2$.
    The solid lines indicate the sum of all components,
    and the dashed, dotted and dot-dashed lines 
    represent the contributions from signal, continuum,
    and $B^+ \to \rho^+\pi^0$.
  }
\end{figure}

The signal yield is extracted using an unbinned maximum-likelihood fit
to the $\mbc - \de$ two-dimensional distribution of the 596 candidates
obtained after all the event selection requirements discussed above.
The fitting function contains components for the signal, 
$B^+ \to \rho^+\pi^0$ and $\qq$ background.
The probability density functions (PDFs) for the signal and for $B^+ \to \rho^+\pi^0$ are taken from
smoothed two-dimensional histograms obtained from large MC samples.
For the signal PDF, discrepancies between the peak positions and resolutions
in data and MC due to imperfect simulation of the $\pi^0$ energy
are calibrated using $\dz\to\pizpiz$ decay.
The invariant mass distribution of $D^0$ are fitted with a bifurcated Gaussian
for data and MC, and
the observed discrepancies in peak position and width are converted to 
the differences of peak position and resolution of $\de$ in our signal PDF,
since observed differences are caused by imperfect simulation of $\pi^0$ energy.
For the $D^0$ daughter particles, similar momentum ranges and the same reconstruction procedures are used as those for the signal daughters.
We find a $3 \pm 9 \ {\rm MeV}$ difference between MC and data for the $\de$ peak position and a 35 $\pm 12$ \% discrepancy in the $\de$ resolution.
To obtain the two-dimensional PDF for the continuum background,
we multiply the $\de$ PDF, which is modeled with a first-order polynomial,
with the $\mbc$ PDF, for which we use the ARGUS function~\cite{argus}.
In the fit, the shapes of the signal and $B^+ \to \rho^+\pi^0$ PDFs are fixed,
the normalization of $B^+ \to \rho^+\pi^0$ is fixed according to 
the recent result from the BaBar collaboration~\cite{babar_rhopi},
and all other fit parameters are allowed to float.
The fit results are shown in Fig.~\ref{fig:fit_result}.

The obtained signal yield is $\YIELDWOSYST$ with a statistical significance
(${\cal S}$) of $\SIGMAWORHOPI$, 
where ${\cal S}$ is defined as 
${\cal S}=\sqrt{-2\ln({\cal L}_0/{\cal L}_{N_s})}$, and
${\cal L}_0$ and ${\cal L}_{N_s}$ denote the maximum likelihoods of the
fits without and with the signal component, respectively.
In order to take the uncertainty in the contribution from 
$B^+ \to \rho^+\pi^0$ into account in the significance calculation,
we repeat the fit after increasing this contribution according to
the error in the measurement of its branching fraction.
In this case we find ${\cal S} = \SIGMA$; 
we interpret this value as the significance of our signal.
We also obtain consistent results when the normalization of
the $B^+ \to \rho^+\pi^0$ component is allowed to float in the fit.

We vary each calibration constant for the signal PDF by $\pm 1 \sigma$ 
and obtain systematic errors from the change in the signal yield.
We also vary the $B^+ \to \rho^+\pi^0$ normalization, as described earlier,
and assign systematic errors accordingly.
Adding these errors in quadrature, we find the yield is $\YIELD$.

In order to obtain the branching fraction, 
we divide the signal yield by the reconstruction efficiency,
measured from MC to be $9.9\%$,
and by the number of $\bb$ pairs.
The trigger efficiency is not corrected since it estimated to be $99\%$ using our trigger simulator.
We consider systematic errors on the reconstruction efficiency 
due to possible differences between data and MC.
We assign a total error of $7\%$ due to $\pi^0$ reconstruction efficiency,
measured by comparing the 
ratio of the yields of the $\eta\to\pizpiz\pi^0$ and 
$\eta\to\gamma\gamma$ decays.
The experimental errors on the branching fractions of these decays~\cite{pdg}
are included in this number.
We compare the performance of the continuum suppression requirement
on a control sample of $\bp\to\dzbpip$ in data and MC; 
a systematic error of $2\%$ is assigned.
The efficiency of the MDLR cut for the MC control sample is close to that for the signal MC.
Finally, we assign a systematic error of $0.5\%$ due to the 
uncertainty in the number of $\bb$ pairs ($152.0 \pm 0.7$) $\times \ 10^6$,
and obtain a branching fraction of 
\begin{displaymath}
{\cal B}(B^0\rightarrow\pi^0\pi^0)={\BR}.
\end{displaymath}
The result is stable for variations of the MDLR cut; for example, if we use 0.925 and 0.9 cut, we obtain
1.5 $\times 10^{-6}$ and 1.8 $\times 10^{-6}$, respectively, which are within the systematic error.

In conclusion, we have measured the branching fraction of $\bz\to\pizpiz$
from a data sample of $\nbb$ million $\bb$ pairs 
collected at the $\Upsilon$(4S) resonance with the Belle experiment. 
We obtain $\YIELD$ signal events with a significance of $\SIGMA$ standard deviations ($\sigma$). 
We measure the branching fraction to be $\BR$.
This result supersedes our previous result.
We find evidence for $\bz\to\pizpiz$ 
at a level above most theoretical predictions~\cite{pi0pi0_predictions} and consistent with the BaBar measurement~\cite{BaBar_new}.
While larger values of the branching fraction for $\bz\to\pizpiz$ enhance the feasibility of the isospin analyses~\cite{isospin}, 
more precise measurements of the branching fraction,
in addition to studies of the CP asymmetries in all $B \to \pi\pi$ decays, 
will be required in order to determine $\phi_2$ from the $\pi\pi$ system.

We wish to thank the KEKB accelerator group for the excellent
operation of the KEKB accelerator.
We acknowledge support from the Ministry of Education,
Culture, Sports, Science, and Technology of Japan
and the Japan Society for the Promotion of Science;
the Australian Research Council
and the Australian Department of Education, Science and Training;
the National Science Foundation of China under contract No.~10175071;
the Department of Science and Technology of India;
the BK21 program of the Ministry of Education of Korea
and the CHEP SRC program of the Korea Science and Engineering Foundation;
the Polish State Committee for Scientific Research
under contract No.~2P03B 01324;
the Ministry of Science and Technology of the Russian Federation;
the Ministry of Education, Science and Sport of the Republic of Slovenia;
the National Science Council and the Ministry of Education of Taiwan;
and the U.S.\ Department of Energy.


\begin{thebibliography}{99}
\bibitem{phi1_belle}
  K.~Abe {\it et al.} (Belle Collaboration),
  Phys. Rev. D {\bf 66}, 071102(R) (2002).
\bibitem{phi1_babar}
  B.~Aubert {\it et al.} (BaBar Collaboration),
  Phys. Rev. Lett. {\bf 89}, 201802 (2002).
\bibitem{km}
  M.~Kobayashi and T.~Maskawa,
  Prog. Theor. Phys. {\bf 49}, 652 (1973).
\bibitem{pdg}
  K.~Hagiwara {\it et al.} (Particle Data Group),
  Phys. Rev. D {\bf 66}, 010001 (2002).
\bibitem{phi2_belle}
  K.~Abe {\it et al.} (Belle Collaboration),
  Phys. Rev. D {\bf 68}, 012001 (2003).
\bibitem{phi2_babar}
  B.~Aubert {\it et al.} (BaBar Collaboration),
  Phys. Rev. Lett. {\bf 89}, 281802 (2002).
\bibitem{isospin}
  M.~Gronau and D.~London,
  Phys. Rev. Lett. {\bf 65}, 3381 (1990);
  M.~Gronau, D.~London, N.~Sinha and R.~Sinha,
  Phys. Lett. B {\bf 514}, 315 (2001).
\bibitem{hh_belle}
  K.~Abe {\it et al.} (Belle Collaboration), BELLE-CONF-0311;
  B.C.K.~Casey {\it et al.} (Belle Collaboration),
  Phys. Rev. D {\bf 66}, 092002 (2002).
\bibitem{hh_babar}
  B.~Aubert {\it et al.} (BaBar Collaboration),
  Phys. Rev. Lett. {\bf 91} 021801 (2003).
\bibitem{hh_cleo}
   D.M.~Asner {\it et al.} (CLEO Collaboration),
   Phys. Rev. D {\bf 65}, 031103(R) (2002).
\bibitem{pi0pi0_predictions}
  M.~Beneke and M.~Neubert, hep-ph/0308039;
  Y.-Y.~Keum and A.I.~Sanda, Phys. Rev. D {\bf 67}, 054009 (2003);
  G.~Kramer and W.F.~Palmer, Phys. Rev. D {\bf 52}, 6411 (1995).
\bibitem{theory}
  C.K.~Chua, W.S.~Hou and K.C.~Yang, Mod.\ Phys.\ Lett.\ A {\bf 18}, 1763 (2003).
\bibitem{BaBar_new}
  B.~Aubert {\it et al.} (BaBar Collaboration),
  hep-ex/0308012.
\bibitem{kekb}
  S.~Kurokawa and E.~Kikutani,
  Nucl. Inst. Meth. A {\bf 499}, 1 (2003).
\bibitem{belle}
  A.~Abashian {\it et al.} (Belle Collaboration),
  Nucl. Inst. Meth. A {\bf 479}, 117 (2002).
\bibitem{geant}
  R.~Brun {\it et al.},
  GEANT 3.21, CERN Report No. DD/EE/84-1 (1987).
\bibitem{fw}
  G.~Fox and S.~Wolfram,
  Phys. Rev. Lett. {\bf 41}, 1581 (1978).
\bibitem{fisher}
  R.A.~Fisher,
  Annals of Eugenics {\bf 7}, 179 (1936).
\bibitem{argus}
  H.~Albrecht {\it et al.} (ARGUS Collaboration),
  Phys. Lett. B {\bf 241}, 278 (1990).
\bibitem{babar_rhopi}
  B.~Aubert {\it et al.} (BaBar Collaboration),
  hep-ex/0307087.
\end{thebibliography}
\end{document}